\documentclass[aps,prb,showpacs,twocolumn,superscriptaddress]{revtex4-2}
\usepackage{amssymb}
\usepackage{graphicx}
\usepackage{appendix}
\usepackage{dcolumn}
\usepackage{bm}
\usepackage{amsmath}
\usepackage{epstopdf}
\usepackage{xcolor}
\usepackage{float}
\usepackage{braket}
\usepackage[colorlinks,urlcolor=blue,anchorcolor=blue,linkcolor=blue,citecolor=blue,breaklinks=true]{hyperref}
\usepackage{multirow}
\usepackage{array}
\usepackage{booktabs}
\usepackage{diagbox}
\usepackage{natbib}

\begin{document}
\title{Square lattice model with staggered magnetic fluxes: zero Chern number topological states and topological flat bands}
\author{Li-Xiang Chen}
\thanks{These authors contributed equally to the work.}
\affiliation{College of Physics Science and Technology, Yangzhou University, Yangzhou 225002, China}
\author{Dong-Hao Guan}
\thanks{These authors contributed equally to the work.}
\affiliation{National Laboratory of Solid State Microstructures and Department of Physics, Nanjing University, Nanjing 210093, China}
\affiliation{College of Physics Science and Technology, Yangzhou University, Yangzhou 225002, China}
\author{Lu Qi}
\email{luqi@yzu.edu.cn}
\affiliation{College of Physics Science and Technology, Yangzhou University, Yangzhou 225002, China}
\author{Xiuyun Zhang}
\affiliation{College of Physics Science and Technology, Yangzhou University, Yangzhou 225002, China}
\author{Ying Han}
\email{hany@yzu.edu.cn}
\affiliation{College of Physics Science and Technology, Yangzhou University, Yangzhou 225002, China}
\author{Ai-Lei He}
\affiliation{College of Physics Science and Technology, Yangzhou University, Yangzhou 225002, China}
\date{\today}

\begin{abstract}
Staggered magnetic fluxes (SMF) play a crucial role in achieving Chern insulators (CIs), by which a series of CI models have been established on various lattices. In addition, SMF induced higher-order topological insulator (HOTI) in a lattice model has been reported. In this work, we propose a square lattice model with SMF. We find intracellular SMF can induce zero-Chern-number topological insulator (ZCNTI) at quarter filling which hosts topologically protected edge states characterized by the quantized polarization, in analogy to the topological state in two dimensional Su-Schrieffer-Hegger model. When lattice dimerization and intracellular SMF are introduced, there exists HOTI state at half filling. Furthermore, this model hosts topological flat band (TFB) by considering the next-nearest-neighbor hoppings. Several fractional Chern insulator states are investigated when hard-core bosons are filled into this TFB model.
\end{abstract}
\maketitle

\section{Introduction}
Exploring and classifying topological phases of matters have been a fascinating topic in condensed matter physics. To date, numerous novel topological insulators are theoretically predicted and experimentally realized~\cite{QHE,Haldane,QSH1,QSH2,reazQSH,TCTI,reazTCTI,ReazQAH,Reali_HM,HOTI1,TI1,TI2}. Among them, Chern insulators (CIs), the realization of quantum Hall states without an external magnetic field, have drawn extensive attention for the fascinating topological nature and considerable application prospects. The first CI model is the well-known Haldane model~\cite{Haldane}, which is established on a honeycomb lattice with staggered magnetic fluxes (SMF). The SMF in Haldane model can open the energy gap at half filling, and give rise to topologically non-trivial bands which are characterized by Chern numbers. Inspired by the sight of adding SMF in Haldane model, other CI models have been proposed on various crystalline lattices~\cite{CB0,LDM,KG0,KG1,Lieb0,Lieb1,Lieb2,Star0,Star1,SQOC0,CB1,KG2,Ruby,TR,SQOC1,KG3,Star2,mapleleaf}, even extended to quasicrystalline and amorphous systems~\cite{QCCI,RN358}. Subsequently, the lowest energy bands of CIs can be flatten by tuning the hopping parameters, $i.e.$, the topological flat bands (TFBs)~\cite{KG2,CB1,Neupert}. TFBs is analogue to the Landau level and fractional Chern insulators (FCIs) have been investigated when interacting particles occupy into TFB models~\cite{TR,Star2,WYFTFB,Sheng1, Neupert,Regnault1,YFWang2,Qi0,Bernevig,YLWu,YLWu0,YLWu3,LiuZ0,LiuZ2,LiuZ3,Ronny0,Ronny1,Ronny2,Ronny3,FCI_reviews,FCI_reviews1,FCI_reviews2}. Although the TFB models have been proposed in some 2D lattice systems, to our knowledge, there are a very few TFB models proposed in square lattice models~\cite{CB1,SQOC1}.

Lattice dimerization can open band gaps as well and leads to zero-Chern-number topological insulators (ZCNTIs)~\cite{SSH1,SSH2,SSH3,SSH4,XHu,Liu1,SSH_flux1,SSH_flux2} which belong to a new class of topological phases instead of the CI phase.  One of the ZCNTI states has been proposed on two dimensional (2D) Su-Schrieffer-Hegger (SSH) model~\cite{Liu1} in which gapless edge state emerges at a quarter filling with open boundary condition. Its topological property can be characterized by the quantized polarization~\cite{Liu1}. We notice that the bulk band gap can not be opened at half filling and there are corner-localized bound states in the bulk continuum in the 2D SSH model~\cite{BenalcazarSSH,BenalcazarSSH1,CALi}. Different from 2D SSH model, both the bulk and edge bands are gapped at half filling in the Benalcazar-Bernevig-Hughes model~\cite{HOTI1,HOTI2} and Kekul{\'e} lattice model~\cite{Liu2}, which the topologically protected states appear around corners. This topological insulator belongs to the higher-order topological insulator (HOTI)~\cite{HOTI1,HOTI2,HOTI3,HOTI4,HOTI5}, an exotic topological insulator which boundary state is with at least two dimension lower than the bulk. This 2D HOTI state can be identified by the quantized quadrupole but its polarization is vanished. Other 2D HOTIs have sequentially proposed in several dimerized systems~\cite{HOTI6,HOTI7,HOTI8,HOTI9,HOTI10,HOTI11}. Additionally, other ZCNTIs and CIs have been studied in the dimerized systems with the SMF. There are several unusual features for these models, such as the special CI state with isolated corner states and metallic near-edge states in the dimerized square lattice~\cite{SSH_flux1}, the coexistence of topologically protected edge and corner states in dimerized square-octagon lattice model~\cite{SSH_flux2} which is characterized by nonzero polarization and quadrupole. More intriguingly, in the ruby lattice model, the SMF can give rise to 2D HOTI state without lattice dimerization~\cite{Guan}.

In this work, we investigate several ZCNTIs in a square lattice model with SMF. When the nearest-neighbor (NN) and intracellular next-nearest-neighbor (NNN) hopping processes are introduced and the SMF threads the intracells at the same time, the ZCNTI emerges at 1/4 filling. This ZCNTI is analogue to the topological state in the 2D SSH model which hosts topologically protected edge states characterized by quantized polarization, however, the present ZCNTI is induced by the SMF instead of lattice dimerization. When we also consider lattice dimerization and intercellular SMF, the HOTI with gapped edge states and robust corner states appear at half filling. Additionally, when the intercellular NNN hoppings are added, a TFB with flatness ratio about 38 is found in a large hopping parameter space. Based on this TFB model and considering hard-core bosons filled, we obtain the $\nu=1/2$ and $\nu=2/3$ FCIs. We also discuss how to realize this model based on the superconducting circuits. Our findings reveal the SMF can give rise to a first-order topological state which is identified by quantized polarization, and provide a TFB model to explore FCIs based on the square lattice.

\section{Model and formulation}
\begin{figure}
\includegraphics[scale=0.4]{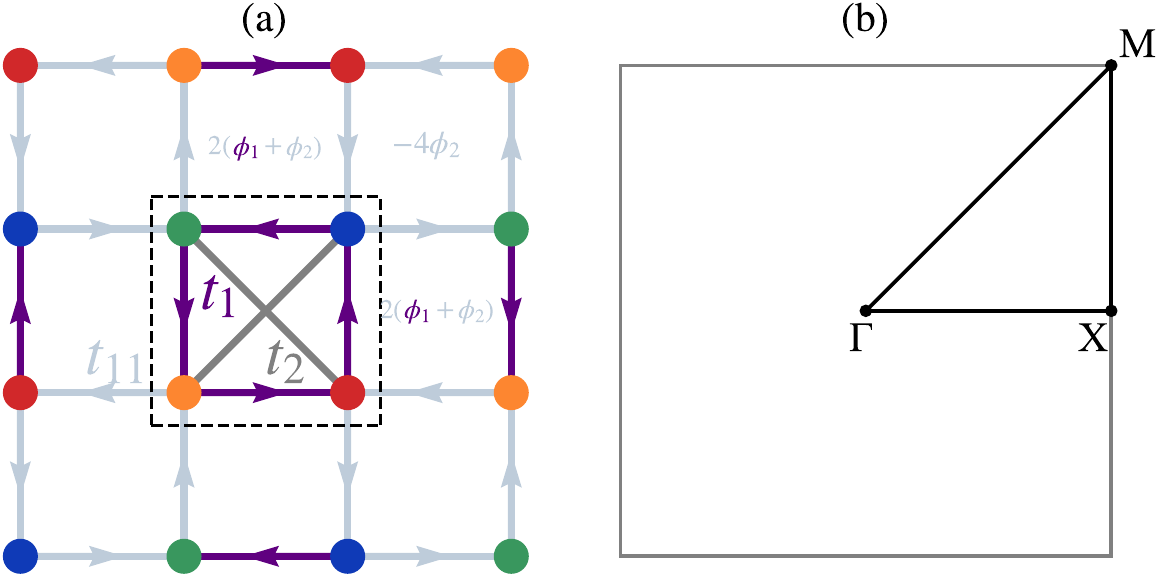}
\caption{(Color online) (a) Schematic structure of an square lattice model with SMF. There are four atoms in a unit cell within a dashed square. The model is spanned by the translation vectors $\vec{a}_1$ and $\vec{a}_2$. Purple and dark gray bonds respectively represent the intracellular NN and NNN hoppings with the corresponding hopping potentials $t_1$ and $t_2$. Light gray bonds are the intercellular hoppings with hopping potential $t_{11}$. The intracellular and intercellular SMF $\phi_1$ and $\phi_2$ are added along the arrows. We also mark the total fluxes in some squares. . (b). The first Brillouin zone of the present model and we highlight higher symmetry points as $\Gamma$ , X and M. }
\label{Model}
\end{figure}

Our model is established on a square lattice by introducing the SMF threading every plaquette as illustrated in Fig.~\ref{Model} (a), and the first Brillouin zone (BZ) of this model is shown in Fig.~\ref{Model} (b). Here, we add the lattice dimerization in this model and consider the intracellular NNN hoppings. The tight-binding Hamiltonian in real space is,
\begin{eqnarray}\label{HAMR}
 \begin{aligned}
H & = t_{1} \sum_{\left\langle\mathbf{r r}^{\prime}\right\rangle}^{}\left[f_{\mathbf{r}^{\prime}}^{\dagger} f_{\mathbf{r}} \exp \left(i \phi_{\mathbf{r}^{\prime} \mathbf{r}}\right)+\text {H.c.}\right]\\
& + t_{2} \sum_{\left\langle\left\langle\mathbf{r r}^{\prime}\right\rangle\right\rangle}\left[f_{\mathbf{r}^{\prime}}^{\dagger} f_{\mathbf{r}}+\text {H.c.}\right]\\
& + t_{11} \sum_{\left\langle\mathbf{r r}^{\prime}\right\rangle^{\prime}}\left[f_{\mathbf{r}^{\prime}}^{\dagger} f_{\mathbf{r}} \exp \left(i \phi_{\mathbf{r}^{\prime} \mathbf{r}}^{\prime}\right)+\text {H.c.}\right],
 \end{aligned}
\end{eqnarray}
where $f_{\mathbf{r}}^{\dagger}(f_{\mathbf{r}})$ creates (annihilates) a spinless fermion at $\mathbf{r}$. $\left \langle ...\right \rangle$ and $\left \langle \left \langle ... \right \rangle \right \rangle $ denote the intracellular NN and NNN pairs of sites and the hopping integrals are $t_{1}$, $t_{2}$, respectively. $\left\langle...\right\rangle^{\prime}$ denotes the intercellular NN pairs of sites, with the intercellular hopping integral $t_{11}$. $\phi_{\mathbf{r}^{\prime} \mathbf{r}}$ is the SMF along the NN hopping processes which the sign indicated by arrows as shown in Fig.~\ref{Model} (a). For the intracellular hoppings, the SMF is $\phi_{\mathbf{r}^{\prime} \mathbf{r}}=\pm \phi _{1}$ and for the intercellular hoppings, the SMF is  $\phi^{\prime}_{\mathbf{r}^{\prime} \mathbf{r}}=\pm \phi _{2}$. One can find that the total flux of this whole lattice model is vanished. Here, we set the NN hopping integral as unit, {\emph {i.e.,}} $t=1.0$.

On account of the SMF,  time-reversal symmetry is broken and may eventually give rise to CIs. To characterize the topological properties, Chern numbers~\cite{TKNN} are calculated by integrating the Berry curvature ${\cal F}$ over the whole BZ, {\emph {i.e.},}
\begin{equation}\label{Ch_num}
{\cal C}_n = \frac{1}{2\pi}\int_{BZ}d^2{\bf k}{\cal  F}_{n}({\bf k}).
\end{equation}
Here, ${\cal C}_n$ and ${\cal  F}_{n}$ respectively denote the Chern number and Berry curvature of the $n$th band defined as ${\cal F}_{n}({\bf k}) = \nabla\times{\cal A}_n({\bf k})$. Here, ${\cal A}_n({\bf k})=-i\langle u_{n{\bf k}}|\nabla_{\bf k}  u_{n{\bf k}} \rangle$ is the Berry connection calculated based on the Bloch wave function of the $n$th band $| u_{n{\bf k}}\rangle$. One can obtain the total Chern number ${\cal C}$ by adding up all Chern numbers of energy bands which is below the chosen fermi energy $E_n$,
\begin{equation}\label{total_Chern}
{\cal C}=\sum_{E_n<E_f} {\cal C}_n.
\end{equation}

In addition to CIs, SMF can lead to 2D HOTIs as well~\cite{Guan}. These HOTIs can be identified on the basis of the polarization and quadrupole moment. We notice that the present model hosts inversion symmetry, and the Hamiltonian satisfies  ${\cal P} H({\bf k}) {\cal P}^{\dagger} = H(-{\bf k})$ where ${\cal P}$ is the inversion operator and $H({\bf k})$ is the Hamiltonian in the momentum space [details shown in the Appendix.~\ref{HAMx}]. With the aid of the inversion symmetry, the polarization and quadrupole moment can be obtained according to the parities~\cite{Liu1,CFang,SSH3,Liu2} at high-symmetric points $\Gamma$, $M$, $X$($Y$) [in Fig.~\ref{Model} (b)],
$\emph i.e.$,
\begin{equation}\label{Quad}
P_{i}=\sum_n p_{i}^{n} \bmod 1, \quad Q_{xy}=\sum_{n} p_{x}^{n} p_{y}^{n},
\end{equation}
and
\begin{equation}\label{Polar}
\quad p_{i}^{n}=\frac{1}{2}\left(q_{i}^{n} \bmod 2\right),\quad(-1)^{q_{i}^{n}}=\frac{\eta^{n}(X_i)}{\eta^{n}(\Gamma)}\quad.
\end{equation}
Here, $X_i$ indicates the high-symmetric point $X$ or $Y$, $\emph i.e.$, $X_1=X$ and $X_2=Y$, and $\eta^n$ is the parity of the $n$th band which is the eigenvalue of inversion operator. Based on the parities, the polarization and quadrupole moment are easily obtained and a detailed proof of Eqs.~(\ref{Quad})-(\ref{Polar}) is presented~\cite{Liu1,CFang,SSH3,Liu2} which the parity of $M$ point is not involved.   Another simple understanding is that, for the polarization ${\bf P}$, ${\bf P}=1/N\sum^{N_{occ}}_{n}\sum_{\bf k} \langle u_{n\bf k}|\hat{\bf x}| u_{n\bf k} \rangle$, it can be written as ${\bf P}=(P_x, P_y)$ where $P_x$ ($P_y$) is along $x$ ($y$) direction. Here, $\hat{\bf x}$ is the position operator, $\hat{\bf x}=(\hat{x},\hat{y})$, and $n$ indicates the occupied energy bands labelled by $n=1,2,...,N_{occ}$. Accordingly, it is related to the parity of $X$ ($Y$) point and the parity of $M$ point is not involved. The eigenvalues of the matrix of inversion operator $\cal P$ is the parity of each high symmetric point. Because of the commutation relation between Hamiltonian $H$ and inversion operator $\cal P$ at high symmetric point, $\emph {i.e.}$, $[{\cal P}, H]=0$, we can obtain the eigenvalues ${\cal P}|u_{n\rm{X}}\rangle = \eta^n(X)|u_{n\rm{X}}\rangle$ where $|u_{n\rm{X}}\rangle$ is the Bloch wave function of the $n$th band at $X$ point and $\eta^n(X)$ corresponds to the parity of the $n$th band at $X$ point.  In general, we choose $q^n_i=0$ or $1$.

\begin{figure}
\includegraphics[scale=0.4]{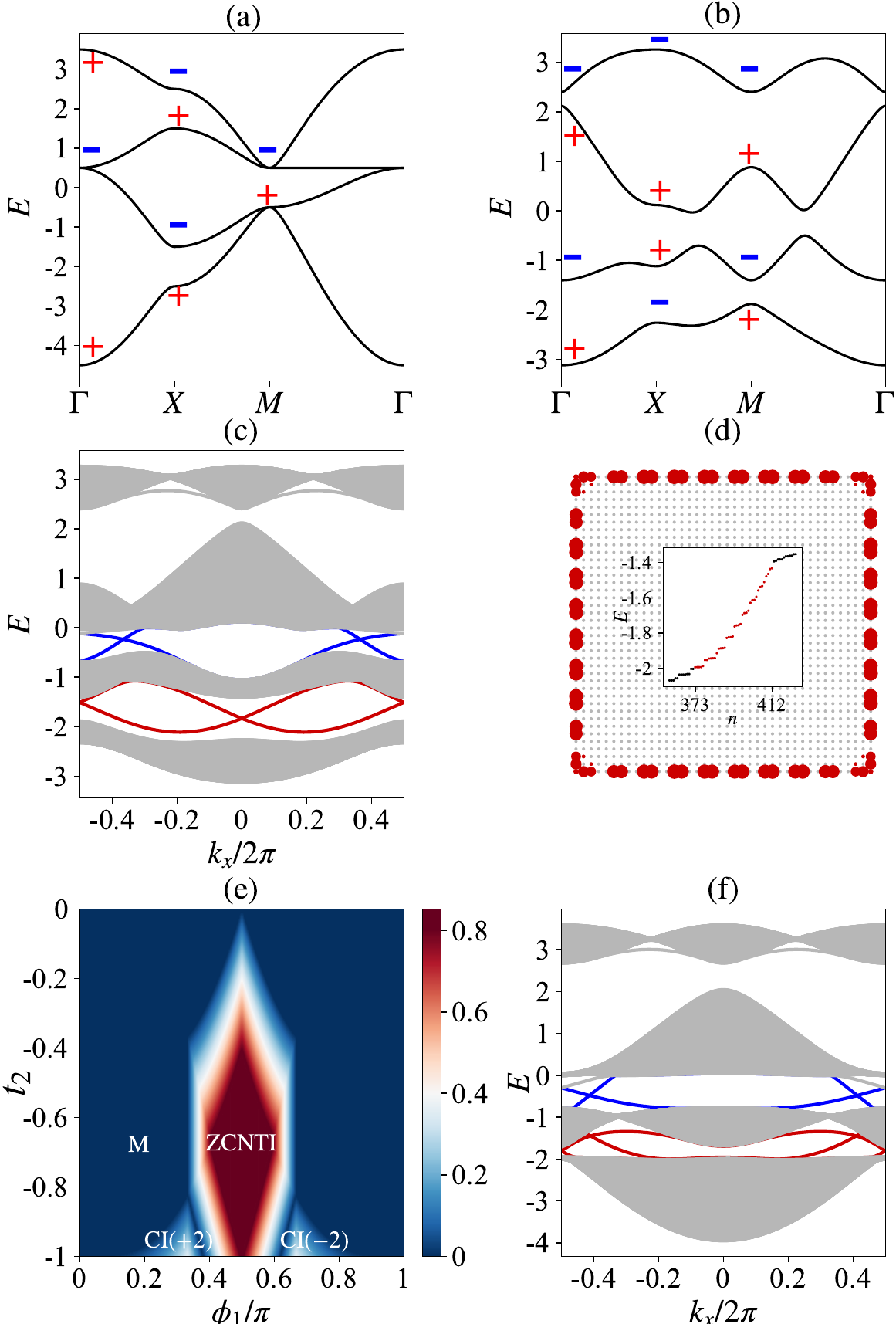}
\caption{(Color online)  Bulk spectra of the model (a) without ($\phi_1= 0.0$) and (b) with SMF ($\phi_1= 0.4\pi$) along intracellular hoppings. ``+" and ``-" indicate the parity at high-symmetric points. (c) Band structure of the model with $\phi_1= 0.4\pi$ on cylinder geometry. Edge states appear and they are marked with red and blue at $1/4$ and half fillings. (d) Real-space density profile of the topological state at $1/4$ filling with open boundary conditions in both directions. The inset displays several bulk (colored with black) and edge (colored with red) states near $1/4$ filling. The chosen lattice hosts $20\times20\times4$ sites. (e) Phase diagram of the present model with the tunable parameters $t_{2}$ and $\phi_1$. This phase diagram contains the metallic (M), zero Chern number topological insulator (ZCNTI), and CI with Chern number two [CI(+2)] and minus two [CI(-2)] phases, respectively. (f) Band structure of the model with $\phi_1= \pi/3$ and $t_2=-0.95$ on cylinder geometry. The color bar denotes the bulk band gap. Here, we set $t_1=t_{11}=1.0$, $t_2=-0.5$ and $\phi_2= 0.0$, unless specified.  }
\label{SMFTI}
\end{figure}

\section{ZCNTI induced by staggered fluxes}
We firstly consider the case where lattice dimerization and SMF are not introduced, $\emph i.e.$, $t_{1}=t_{11}$ and $\phi_1=\phi_2=0$, and the intracellular NNN hopping is added with hopping integral $t_2=-0.5$.  By the Fourier transformation and numerical diagonalization of the Hamiltonian $H(\bf{k})$ on torus geometry, the bulk energy spectrum in the first BZ is shown in Fig.~\ref{SMFTI} (a). There are four bands and the energy bands cross each other at $\Gamma$ and $M$ points which indicates the metallic state without lattice dimerization and SMF for this model, even though the intracellular NNN hopping is introduced. Subsequently, only adding the intracellular SMF, for example, $t_{1}=t_{11}=1.0$, $\phi_1=0.4\pi$ and $\phi_2=0$, the bulk bands open gaps both at quarter and half fillings [details in Fig.~\ref{SMFTI} (b)], suggesting the appearance of insulator states. To identify these insulator phases, we calculate the Chern number of each band and find ${\cal C}_1=0$ and ${\cal C}_2=2$. Notably, the insulator state at half filling belongs to the CI with Chern number two and the corresponding energy band on cylinder geometry (one direction with periodic boundary condition and another direction with open boundary condition) is shown in Fig.~\ref{SMFTI} (c) which hosts two branches of edge states with the same propagating direction. One can find the emergence of edge states at quarter filling as well [colored with red in Fig.~\ref{SMFTI} (c)]. However, we cannot directly diagnose the topological properties of this state only based on the Chern number and edge states.

To identify the insulator state at quarter filling, we require to calculate the polarization and quadrupole moment. Here, we obtain the parities at high-symmetric points both of the metallic and insulator states and mark the signs of parities in the corresponding energy bands [in Figs.~\ref{SMFTI} (a) and (b)]. Compared to the parities at $\Gamma$ and $X$ points of the lowest band in Fig.~\ref{SMFTI} (a), we find the sign of parity at $\Gamma$ point is still ``+", however, the sign of parity at $X$  point changes from ``+" to ``-" when the intracellular SMF is introduced. This change attributes to the closing and reopening the bulk band gap. Similarly, the sign of parity at $Y$  point is the same as the $X$ point because of the four-fold rotational symmetry.  The polarization is obtained based on Eqs.~(\ref{Quad}) and~(\ref{Polar}), ${\emph {i.e.}}$, $P_x=P_y=1/2$ [$\eta^1(\Gamma)=+1$, $\eta^1(X)=\eta^1(Y)=-1$, and $q^1_x=q^1_y=1$, see Fig.~\ref{SMFTI} (b)]. The corresponding edge state is gapless [in Fig.~\ref{SMFTI} (c)] and its real-space density distribution is shown in Fig.~\ref{SMFTI} (d). No robust corner state exists based on the density distribution of boundary state [Fig.~\ref{SMFTI} (d)] though this topological state hosts quantized quadrupole moment ($ Q_{xy}= 1/2\times 1/2 = 1/4$), which manifests this topological state belongs to a ZCNTI instead of HOTI.

This ZCNTI shares the similar topological properties of the topological state with zero Berry curvature in 2D SSH model~\cite{Liu1} whose polarization is quantized to $1/2$ as well. One can find there is no band gap closed when the ZCNTI evolves to the topological state with vanished Berry curvature (details in the Appendix.~\ref{ZCNTIdets}) and their topological origin is consistent. However, for these two states, there are two slight differences: i).The distribution of Berry curvature of this ZCNTI is non-zero everywhere because of the broken time-reversal symmetry and ii). the present ZCNTI is induced by the SMF, instead of the lattice dimerization~\cite{Liu1}. By tuning the hopping parameters, the topological phase transitions are explored [details in Fig.~\ref{SMFTI} (e)]. When $t_2>-0.8$, SMF can directly open the band gap and gives rise to the ZCNTI. When $t_2$ is near to $-1$, CIs with Chern number $|{\cal C}_1|=2$ appear and the corresponding energy bands on cylinder geometry is shown in Fig.~\ref{SMFTI} (f) where two branches of edge states appear at $1/4$ filling. Based on the phase diagram [Fig.~\ref{SMFTI} (e)], the CI phase can transition to ZCNTI phase by tuning the SMF. As a consequence, the SMF along the intracellular hoppings ($\phi_1$) can lead to ZCNTI state at quarter filling in this square lattice model with intracellular NNN hoppings.

\section{HOTI at half filling}
\begin{figure}
\includegraphics[scale=0.38]{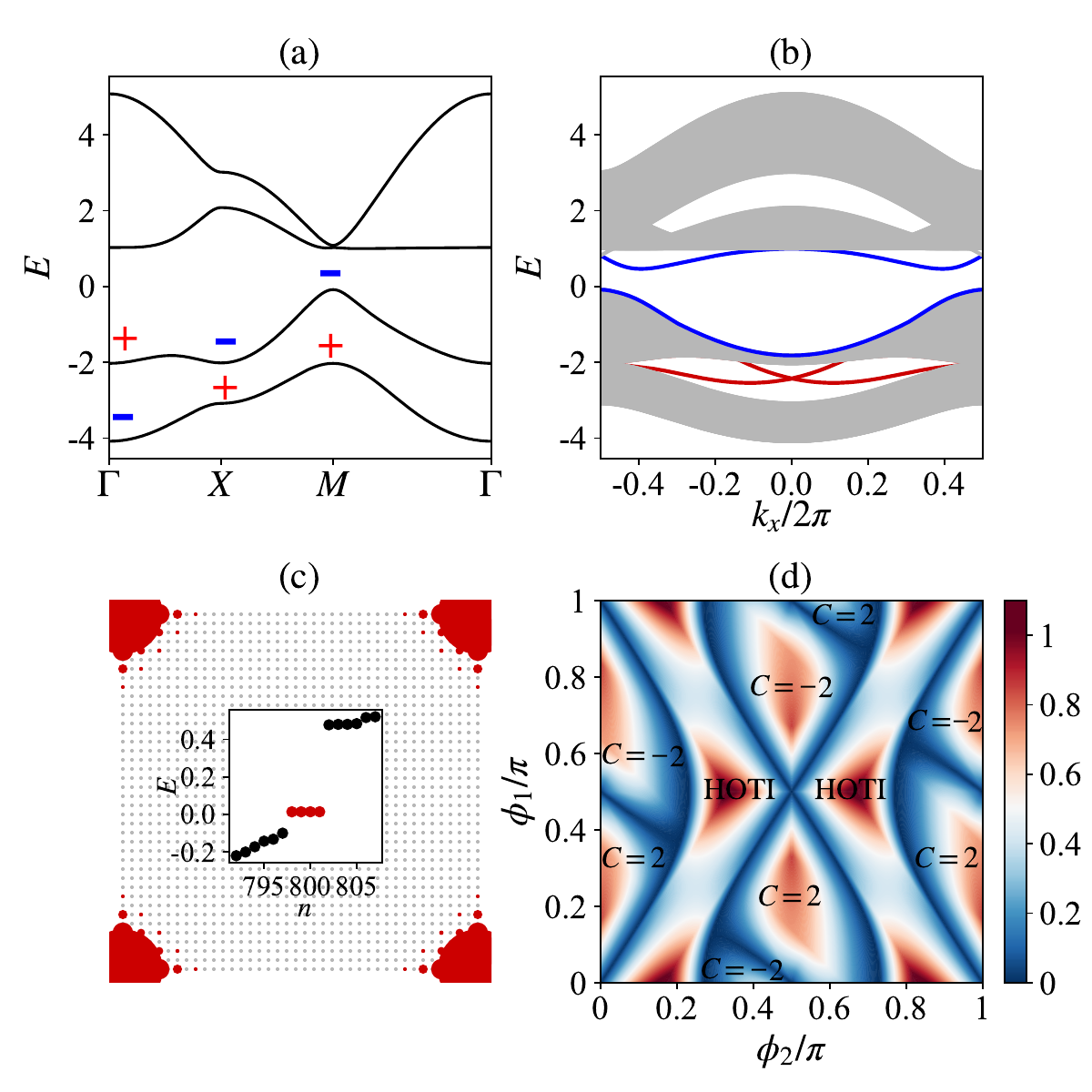}
\caption{(Color online) (a) Bulk spectra of the square lattice model with both intracellular and intercellular SMF ($\phi_1$ and $\phi_2$). Here we choose $\phi_{1}= 0.5\pi$, $\phi_{2}= 0.33\pi$. ``+" and ``-" indicate the parity at high-symmetric points. (b) Energy bands on cylinder geometry. The edge states are gapped at half filling. (c) Density distribution of the HOTI state on a lattice with $20 \times 20 \times 4$ points with open boundary conditions along both $x$ and $y$ directions. Corner states emerge marked with red which reveals the existence of HOTI. (d) Phase diagram at half filling with tunable $\phi_1$ and $\phi_2$. Here, we set $t_{11}=1.5$ and $t_2=-0.5$. The color bar denotes the bulk band gap. Several phases appear, including the CI phase with $\mathcal{C}= \pm 2$ and two HOTIs.}
\label{HOTIx}
\end{figure}

A ${\cal C}=2$ CI state at half filling is obtained in the square model with the intracellular NN and NNN hoppings. Even though the lattice dimeriztion is introduced, ${\cal C}=2$ CI state still exists and no phase transition occurs by increasing the intercellular hopping potential $t_{11}$~\cite{SSH_flux1}. For this ${\cal C}=2$ CI state, there exists special boundary states around the corner and near-edge with large lattice dimerization~\cite{SSH_flux1}. Here, we continue to introduce the intercellular SMF ($\phi_2$) and by increasing value of $\phi_2$, the bulk gap closes and reopens at half filling. For example, when $t_{11}=1.5$, $t_2=-0.5$, $\phi_1=0.5\pi$ and $\phi_2=0.33\pi$, we find the bulk band gap opened at half filling [in Fig.~\ref{HOTIx} (a)] and the total Chern number of this state is vanished which is different from the special ${\cal C}=2$ CI state~\cite{SSH_flux1}.

To identify the topological properties of the insulator state with vanished Chern number at half filling, the polarization and quadrupole moment are calculated based on the parities and marked in Fig.~\ref{HOTIx} (a). Here, $\eta^1(\Gamma)=-1$, $\eta^1(X)=\eta^1(Y)=+1$, $q^1_x=q^1_y=1$ and $\eta^2(\Gamma)=+1$, $\eta^2(X)=\eta^2(Y)=-1$, $q^2_x=q^2_y=1$. One can easily find the vanished polarization ($P_x=P_y=0$) and quantized quadrupole moment ($Q_{xy}=1/2$) at half filling based on Eqs.~(\ref{Quad}) and~(\ref{Polar}), which manifests the existence of HOTI state. We also investigate the energy bands on cylinder geometry and there are gapped edge states at half filling [Fig.~\ref{HOTIx} (b)], which is one feature of HOTI. We numerically diagonalize the tight-binding Hamiltonian with open boundary condition and find that four-fold degenerate states appear between the edge and bulk bands. Simultaneously, we find that these four-fold degenerate states belong to corner state and the density distribution of this state is shown in Fig.~\ref{HOTIx} (c) where most of the density gathers around the corners. Subsequently, we present the phase diagram [shown in Fig.~\ref{HOTIx} (d)] by tuning the intracellular and intercellular SMF ($\phi_1$ and $\phi_2$). When $\phi_2=0$, this model supports ${\cal C}=\pm 2$ CI phases. When $\phi_2$ is introduced, HOTI phase appears [also see the Appendix.~\ref{ZCNTIdets}]. This HOTI seems to be originated from the intercellular SMF, however, the HOTI state is vanished when the model is without lattice dimerization, even though the intercellular SMF is added. Based on the phase diagram [in Fig.~\ref{HOTIx} (d)], we find ${\cal C}=\pm2$ CI phase transitions to the HOTI phase by tuning the intracellular and intercellular SMF at half filling.

\begin{figure}
\includegraphics[scale=0.45]{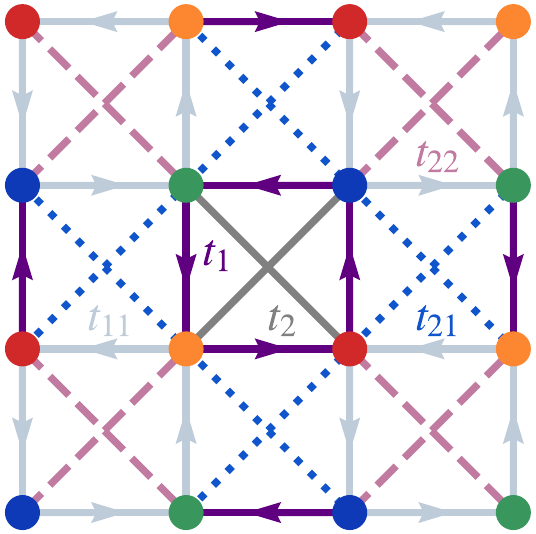}
\caption{(Color online) Schematic structure of the model by adding the intercellular NNN hoppings with blue and red dashed lines. The corresponding hopping potentials are $t_{21}$ and $t_{22}$.}
\label{LM0}
\end{figure}

\section{TFB model and FCIs}
\begin{figure}[!htb]
\includegraphics[scale=0.45]{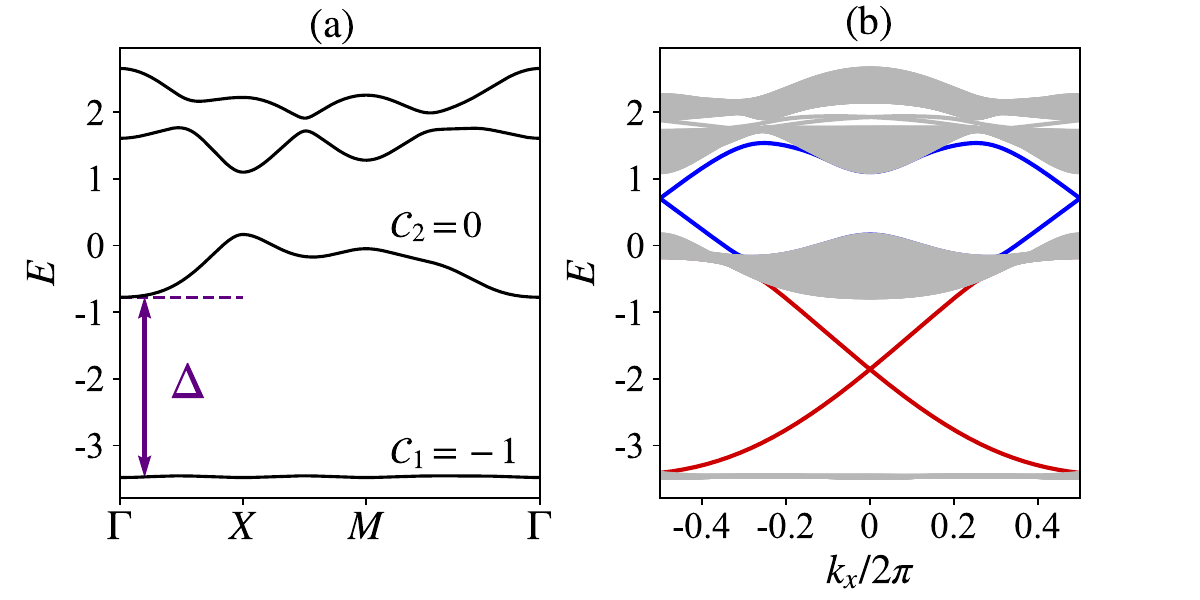}
\caption{(Color online) TFB model in square lattice. (a). Bulk energy band of the TFB model on torus geometry. The lowest energy band is very flat and it separate from other energy bands with a large gap $\Delta$. Chern number of this TFB is ${\cal C}=-1$ which (b) a robust edge state (colored with red) appear connecting the lowest and the second lowest bulk bands.}
\label{TFBM}
\end{figure}

Several TFB models have been proposed on various lattice models, nevertheless, a very few TFB models are constructed on the square lattice~\cite{CB1,SQOC1}. Here, we add the intercellular NNN hopping processes for this model and the corresponding hoppings are $t_{21}$ and $t_{22}$ (the details of hopping processes shown in the Fig.~\ref{LM0}). We search the TFB in a large mount of hopping parameter spaces $\{t_{11},t_2,t_{21},t_{22},\phi_1\,\phi_2\}$. A very flat energy band with flatness ratio more than 38 is found [in Fig.~\ref{TFBM} (a)] and the corresponding hopping parameters are $\{t_{11},t_2,t_{21},t_{22},\phi_1,\phi_2\}= \{-0.75, 0.8,  -0.65, -0.4, -0.67\pi,0\}$. The flatness ratio is defined as $\Delta/W$, where $\Delta$ is the band gap between the first and second lowest energy bands and $W$ denotes the band width of the lowest energy band. The Chern number of this flat band can be calculated based on Eq.~(\ref{Ch_num}) and its Chern number is ${\cal C}=1$, which reveals that this flat band belongs to a TFB. Energy band of this TFB model on cylinder geometry is displayed in Fig.~\ref{TFBM} (b) and one branch of chiral edge state connecting the first and the second lowest bulk bands appears, corresponding to the ${\cal C}=1$ CI state.

\begin{figure}[!htb]
\includegraphics[scale=0.65]{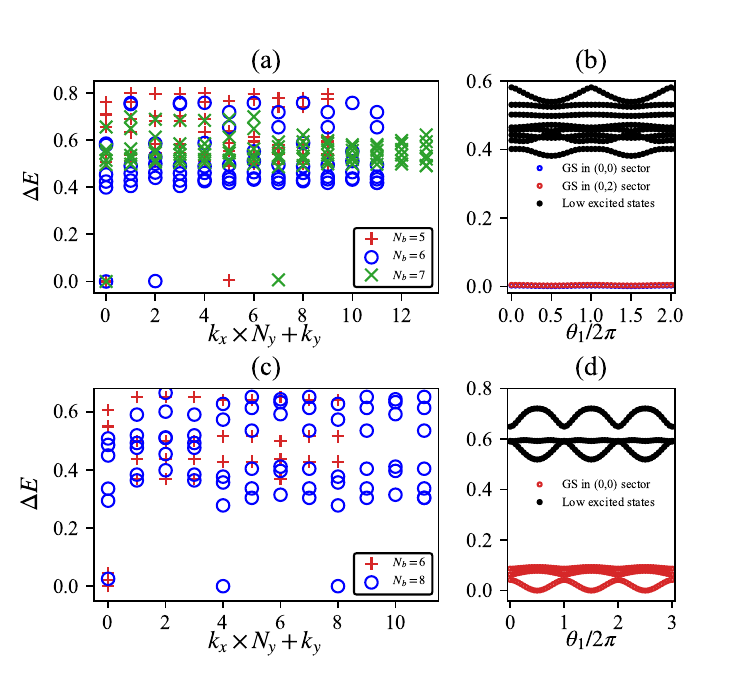}
\caption{(color online).FCI states in the TFB model filling with hard-core bosons. (a) Low-energy spectrum $\Delta E = E_n-E_0$ of $\nu=1/2$ FCI versus the momentum $k_x\times N_y+k_y$. (b) Low-energy spectrum $\Delta E$ of $\nu=1/2$ FCI versus $\theta_1$ with a fixed $\theta_2=0$. Two-fold quasidegenerate ground states (GS) are in momentum sectors $(k_x,k_y)=(0,0)$ and $(k_x,k_y)=(0,2)$. (c) Low-energy spectrum $\Delta E$ of $\nu=2/3$ FCI versus the momentum $k_x\times N_y+k_y$. (d) Low-energy spectrum $\Delta E$ of $\nu=2/3$ FCI versus $\theta_1$ with a fixed $\theta_2=0$. Three-fold quasidegenerate GS are in the momentum sector $(k_x,k_y)=(0,0)$. }
\label{FCIx}
\end{figure}

For the TFB model with a large flatness ratio, FCIs can be explored when hard-core bosons fill into this TFB model. Here, we adopt the real-space exact diagonalization method without any band projection. For the ED study, $N_b$ hard-core bosons fill into a $N_x \times N_y$ unit cells with total $4\times N_x \times N_y$ sites on torus geometry. The filling factor is $\nu=N_b/(N_x \times N_y)$ because only the lowest energy band (corresponding to the TFB) is filled with hard-core bosons. Thanks to the translational symmetry, low-energy  spectrum can be obtained by diagonalizing the many-body Hamiltonian in each  momentum sector ${\bf q}=(2\pi k_x/N_x, 2\pi k_y/N_y)$ which are integer quantum numbers. Dimensions of Hilbert space for FCIs with various lattice sizes are shown in the Appendix.~\ref{dimFCIs}.  We consider $N_b=5$, $N_b=6$ and $N_b=7$ bosons filling in the TFB model with $40-$, $48-$ and $56-$sites, respectively. Based on the exact diagonalization results, two-fold quasidegenerate ground states are obtained and these ground states are well separated from the higher energy spectrum with large gaps [shown in Fig.~\ref{FCIx} (a)]. Simultaneously, these quasidegenerate ground states can evolve into each other with level crossings and they are still separated from other excited energy spectrum with large gaps when tuning the boundary phases [shown in Fig.~\ref{FCIx} (b)]. These results reveal the existence of $\nu=1/2$ FCI state. We also consider $N_b=6$ and $N_b=8$ bosons filling in the TFB model with $36-$ and $48-$sites and three-fold quasidegenerate ground states are observed [displayed in Fig.~\ref{FCIx} (c)]. These quasidegenerate ground states are separated from other excited energy spectrum with large gaps and they can evolve into each other with level crossings [in Fig.~\ref{FCIx} (d)], which manifests the existence of $\nu=2/3$ FCI state.

\section{Superconducting-circuit realization of the model}
\begin{figure}[!htb]
	\centering
	\includegraphics[width=1.0\linewidth]{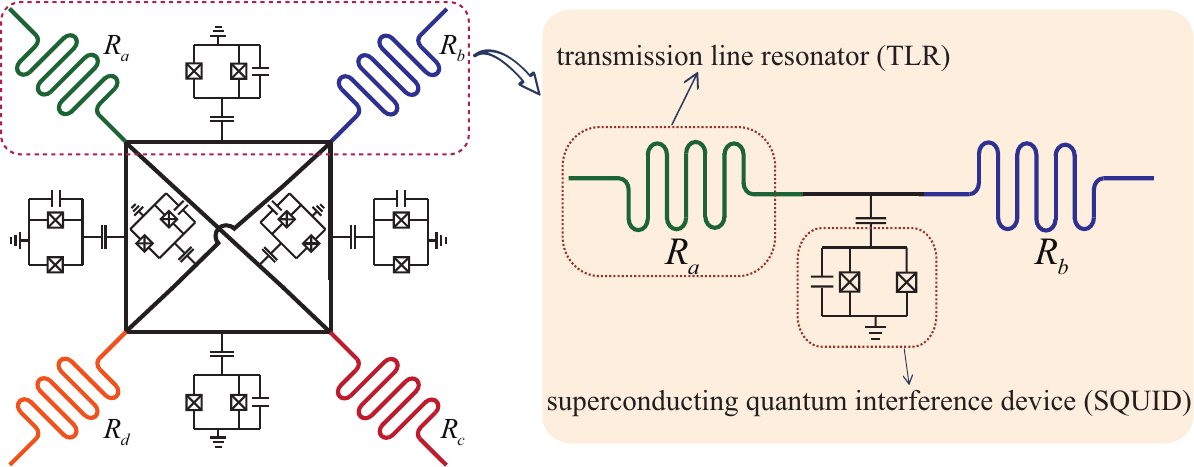}
	\caption{Schematic of the transmission line resonator (TLR) square lattice simulator. The four sites in one unit cell of square lattice are mapped by four TLRs ($R_{a}$, $R_{b}$, $R_{c}$, and $R_{d}$). Two TLRs are coupled with each other by a superconducting quantum interference device (SQUID), leading the coupling between two TLRs can be modulated with time.}
	\label{fig6}
\end{figure}

We have demonstrated the existence of ZCNTI states and TFBs in this square lattice model. In this sector, we briefly discuss how to realize this model with SMFs and NNN hoppings with the aid of superconducting circuits~\cite{SCu1,SCu2,SCu3,SCu4,SCu5,SQC01,SQC00}. As shown in Fig.~\ref{fig6}, the square lattice with SMF along the NN bonds and the NNN hoppings can be mapped by four transmission line resonators (TLRs) $R_{a}$, $R_{b}$, $R_{c}$, and $R_{d}$ where two resonators are coupled with each other by a superconducting quantum interference device (SQUID). In this way, the coupling between two resonators can be modulated dynamically. For example, for the resonators $R_{a}$ and $R_{b}$, the Hamiltonian can be written as $H=H_{0}+H_{\mathrm{hopping}}$ where
\begin{eqnarray}\label{e01}
H_{0}&=&\omega_{a}a^{\dagger}a+\omega_{b}b^{\dagger}b,
\end{eqnarray}
and
\begin{eqnarray}\label{e02}
H_{\mathrm{hopping}}&=&g(t)(a^{\dagger}+a)(b^{\dagger}+b),
\end{eqnarray}
being the free energy of two resonators and coupling between two resonators respectively. Here $\omega_{j}$ ($j=a,b$) denotes the eigenfrequencies of the two resonators and $g(t)=2J\cos[(\omega_{a}-\omega_{b})t-\theta]=J\{e^{i[(\omega_{a}-\omega_{b})t-\theta]}+\mathrm{h.c.}\}$ represents the modulated coupling amplitude induced by the SQUID between two resonators. Under the rotation frame with respect to the free energy $H_{0}$, the effective hopping Hamiltonian under the assumption of $\omega_{a}+\omega_{b}\gg|\omega_{a}-\omega_{b}|$ can be written as
\begin{eqnarray}\label{e03}
H_{\mathrm{eff}}^{\prime}&=&g(t)[e^{i(\omega_{a}-\omega_{b})t}a^{\dagger}b+\mathrm{H.c.}].
\end{eqnarray}
Note that, if we further take the assumption of $|\omega_{a}-\omega_{b}|\gg J$, the final effective Hamiltonian can be written as
\begin{eqnarray}\label{e04}
H_{\mathrm{eff}}&=&Je^{i\theta}a^{\dagger}b+\mathrm{H.c.} .
\end{eqnarray}
In this way, the phase in the effective hopping can be obtained by the SQUID. And the NNN hoppings can be introduced by the SQUID as well.

More specifically, we usually use the rotating wave approximation (RAW) to ignore the high-frequency oscillation terms in the effective dynamic of Hamiltonian, in which the RAW is reasonable only when the driving amplitude is much smaller than the driving frequency~\cite{SQC00}. Thus, we take the condition of weak coupling in the derivation of effective Hamiltonian. Furthermore, we take the effective Hamiltonian by adding the appropriate driving which we can implement the detection of some topological characteristics~\cite{SQC01,SCu5}. Specifically, when the superconducting circuit system has the driving and decay simultaneously, the system may enter into the steady state under the long time limit. After that, the topological state can be detected by the statistical characteristic of mean value of cavity field. Here we stress that the mapping between the effective Hamiltonian and the topological tight binding model has been investigated widely~\cite{SQC01,SCu5,SQC00}, which ensures the rationality of this method.

Apart from superconducting circuits, some promising approaches to realize CIs have been proposed as well, such as the ultracold atomic~\cite{Reali_HM,EOL,EOL1}, the photonic~\cite{PCCI0,PCCI1,PCCI2}, the phononic~\cite{PhonoCI} and electrical-circuit systems~\cite{CirCI1}. Notably, CIs have been experimentally achieved in Moir{\'e} systems~\cite{MIRQAH1,MIRQAH2,MIRQAH3}. Accordingly, twisted systems provide a possibility to realize the present model and corresponding topological states.

\section{Summary}
Several ZCNTIs are investigated in the square lattice model with SMF. At $1/4$ filling, we find first-order ZCNTI with quantized polarization. This ZCNTI is analogue to the topological state with zero Berry curvature reported in 2D SSH model. Different from the topological insulator in 2D SSH model, the present ZCNTI is induced by the intracellular SMF instead of lattice dimerization. At $1/2$  filling, HOTI can be found when both lattice dimerization and intercellular SMF are introduced. This HOTI state hosts gapped edge states and robust corner states which can be characterized by the quantized quadrupole moment. When NNN hopping processes are considered, a TFB with flatness ratio about 38 is obtained. We further reveal the existence of $\nu=1/2$ and $\nu=2/3$ FCI states based on this TFB model filling with hard-core bosons.
Our findings suggest that the SMF can give rise to the first-order topological states with vanished Chern number and provide a TFB model on the square lattice to explore FCIs.

\section*{Acknowledgments}
This work is supported by the National Natural Science Foundation of China (Grant Nos. 12204404, 12304557),
and the Natural Science Foundation of Jiangsu Higher Education Institutions of China (Grant No. 24KJB140022).

\section*{APPENDIX}
\setcounter{equation}{0}
\renewcommand \theequation{A.\arabic{equation}}
\setcounter{section}{0}
\renewcommand \thesection{A\arabic{section}}

\begin{figure}
\includegraphics[scale=0.4]{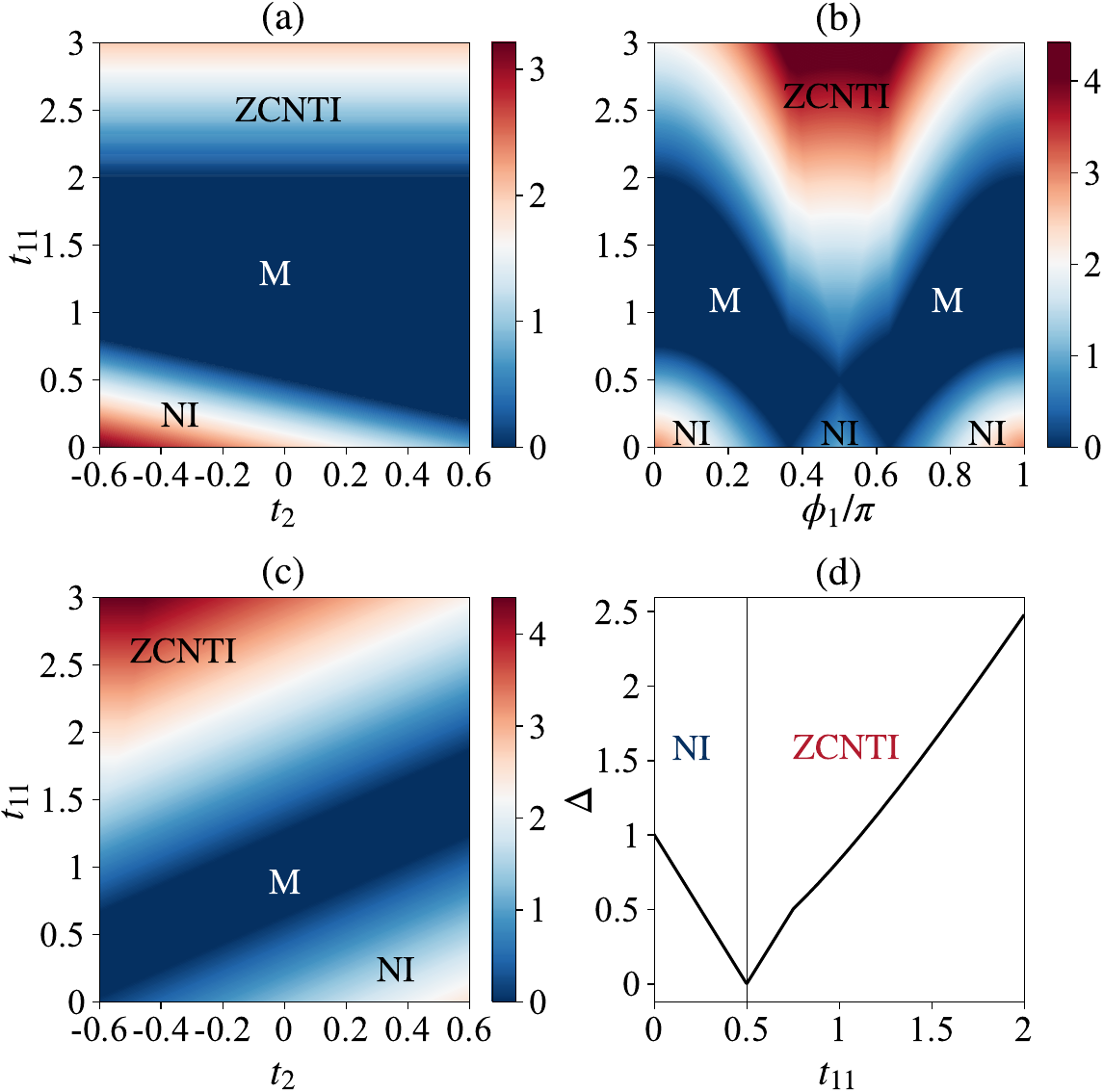}
\caption{(Color online) (a)-(c) The phase diagrams of the square lattice model at $1/4$ filling by tuning the hopping parameters. This model supports the metallic (M), the normal insulator (NI) and the zero Chern number topological insulator (ZCNTI) phases. (a). $\phi_1=\phi_2=0$, (b) $t_2=-0.5$ and $\phi_2=0$, and (c) $\phi_1=0.4\pi$ and $\phi_2=0$. (d) Band gap $\Delta$ versus $t_{11}$ and we set $t_2=-0.5$, $\phi_1=0.4\pi$ and $\phi_2=0$. }
\label{A2x}
\end{figure}
\section{HAMILTON IN MOMENTUM SPACE and INVERSION OPERATOR} \label{HAMx}
The Hamiltonian matrix in the momentum space can be obtained by Fourier transformation of the real-space Hamiltonian [see Eq.~(\ref{HAMR})], {\emph i.e.},
\begin{equation}\label{Hkx}
H\left( \mathbf{k} \right) =\left( \begin{matrix}
	H_{11}&		H_{12}&		H_{13}&		H_{14}\\
	H_{21}&		H_{22}&		H_{23}&		H_{24}\\
	H_{31}&		H_{32}&		H_{33}&		H_{34}\\
	H_{41}&		H_{42}&		H_{43}&		H_{44}\\
\end{matrix} \right),
\end{equation}
where $H_{12}=H_{21}^{*}=t_1e^{-i\phi _1}+t_{11}e^{-ik_x}e^{-i\phi _2}$, $H_{13}=H_{31}^{*}=t_2$, $H_{14}=H_{41}^{*}=t_1e^{i\phi _1}+t_{11}e^{i\phi _2}e^{-ik_y}$, $H_{23}=H_{32}^{*}=t_1e^{-i\phi _1}+t_{11}e^{-i\phi _2}e^{-ik_y}$, $H_{24}=H_{42}^{*}=t_2$, and $H_{34}=H_{43}^{*}=t_1e^{-i\phi _1}+t_{11}e^{-i\phi _2}e^{ik_x}
$. Here, we consider the intracellular NN and NNN hopping processes and the intercellular hoppings. For this case, there is no TFB in this model. To obtain a TFB model, we add the intercellular NNN hopping processes, {\emph i.e.}, $H_{13}=H_{31}^{*}=t_2+t_{21}e^{-ik_x}+t_{21}e^{-ik_y}+t_{22}e^{-ik_x-ik_y}
$ and $H_{24}=H_{42}^{*}=t_2+t_{21}e^{ik_x}+t_{21}e^{-ik_y}+t_{22}e^{ik_x-ik_y}$. The inversion operator is
\begin{equation}\label{Invson}
{\cal P}=\left( \begin{matrix}
	0&		0&		1&		0\\
	0&		0&		0&		1\\
	1&		0&		0&		0\\
	0&		1&		0&		0\\
\end{matrix} \right).
\end{equation}
One can easily find ${\cal P} H({\bf k}) {\cal P}^{\dagger}=H(-{\bf k})$.

\begin{figure}
\includegraphics[scale=0.4]{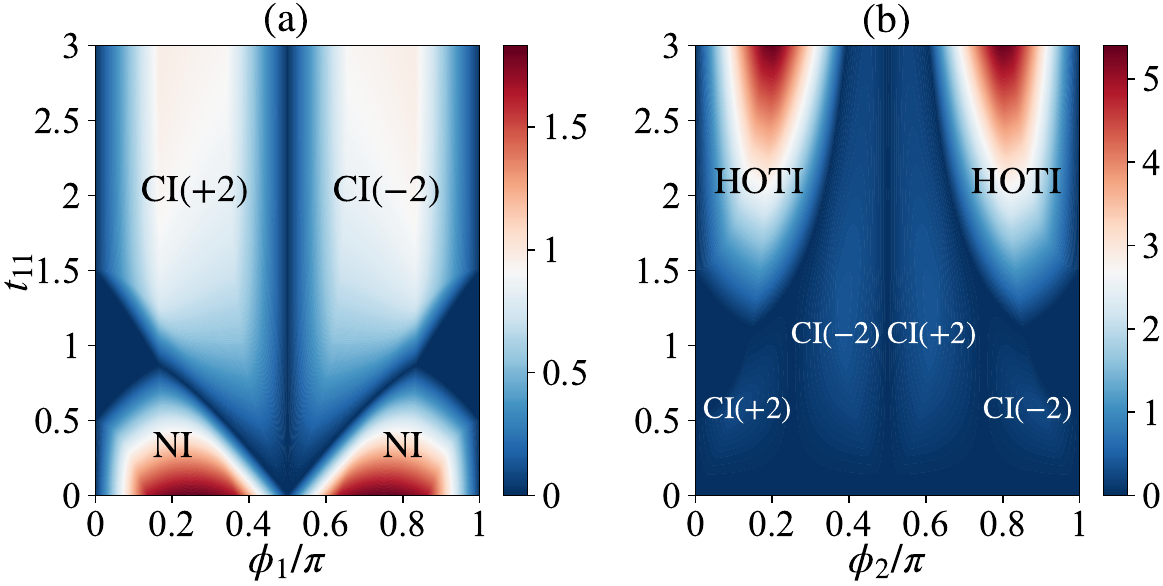}
\caption{(Color online) Phase diagrams of the present model at half filling. (a) Only by adding the intracellular SMF $\phi_1$, the model supports normal insulator (NI) and $|{\cal C}|=2$ Chern insulator (CI) phases. (b). Only by adding the intercellular SMF $\phi_2$, this model supports $|{\cal C}|=2$ CI phases and higher-order topological insulator (HOTI) phases. Here, we set $t_2=-0.5$.}
\label{A3x}
\end{figure}

\section{Details of Zero Chern number topological phases} \label{ZCNTIdets}
For the ZCNTI state induced by the SMF at 1/4 filling, topologically protected edge state appears and no corner states emerge, which suggests this ZCNTI belongs to the first-order topological state. We find this ZCNTI and topological state with zero Berry curvature in the 2D SSH model belong to the same topological phase. We first present the phase diagram of 2D SSH model with the intracellular hoppings with hopping integral $t_2$ at $1/4$ filling in Fig.~\ref{A2x} (a). When the hopping potential of lattice dimerization is larger than 2.0 ($t_{11}>2.0$), the phase transitions from metal to the ZCNTI and $t_2$ hardly change this topological phase region [details in Fig.~\ref{A2x} (a)]. When we choose a fixed $t_2$ ($t_2=-0.5$), a phase diagram can be present by tuning the lattice dimerization ($t_{11}$) and the intracellular SMF ($\phi_1$) [in Fig.~\ref{A2x} (b)]. Notably, the phase with $t_{11}=1.0$ and $\phi_1=0.4\pi$ and the phase with $t_{11}=3.0$ and $\phi_1=0.0\pi$ belong to the ZCNTI phase because of no bulk band gap closed and reopened. Accordingly, the present ZCNTI and zero Berry curvature topological phase belong to the same phase. Additionally, we present the phase diagram by tuning $t_{11}$ and $t_2$ with the fixed $\phi_1=0.4\pi$ in Fig.~\ref{A2x} (c) and the SMF induced ZCNTI phase has the same topological properties as the case with a strong lattice dimerization. We also explore the phase transition by tuning the hopping of lattice dimerization ($t_{11}$) with $t_2=-0.5$, $\phi_1=0.4\pi$ and $\phi_2=0$ [in Fig.~\ref{A2x} (d)]. The phase transition occurs at $t_{11}=0.5$.

Another ZCNTI phase is found at half filling which the intercellular SMF ($\phi_2$) in introduced. When only the intracellular SMF, this model only supports normal insulator and $|{\cal C}|=2$ CI phases, even though with a strong lattice dimerization [in Fig.~\ref{A3x} (a)]. By adding the intercellular SMF and with a large lattice dimerization, HOTI phase appear. Here, we present the phase diagram of this model by tuning $t_{11}$ and $\phi_2$ without the intracellular SMF [in Fig.~\ref{A3x} (b)]. When the lattice dimerization is large enough, HOTI phase can appear. Consequently, the lattice dimerization and the intercellular SMF give rise to the HOTI.

\section{Dimensions of Hilbert space for many-body Hamiltonian} \label{dimFCIs}
To obtain FCIs, we use real space diagonalization method to calculate the low energy spectrum. Because of the existence of translational symmetries along $x-$ and $y-$ directions, we can implement the exact diagonalization in the subspace of the Hilbert space with the dimension reduced by a factor $1/(N_xN_y)$. We consider $N_b$ particles fill into a  $N_{tot} = 4 \times N_x \times N_y-$site model, the dimension of Hilbert space is $\frac{N_{tot}!}{N_b!(N_{tot}-N_b)!} $ and the dimension of subspace is about $\frac{N_{tot}!}{N_b!(N_{tot}-N_b)!} \times \frac{1}{(N_xN_y)}$. For example, if we consider $N_b=7$ hard-core bosons fill into a 56-site TFB model, the dimension of subspace in each momentum sector is about $1.6\times10^7$. The low-energy spectrum of the Hamiltonian with dimension about $1.6\times10^7$ can be obtained based on the Lanczos algorithm. We provide the table to show the dimensions of Hilbert space for the TFB model with different system sizes and various filling numbers in the Table~\ref{tab1}.

\begin{table}
\begin{tabular}{c| c| c| c}
\hline\hline
  $\nu=1/2$    &  $N_s=40$, $N_b=5$ &  $N_s=48$, $N_b=6$ &  $N_s=56$, $N_b=7$  \\
\hline
  $\cal D$   &    $\sim 65800$    &   $\sim 1022626$     & $\sim 16565528$    \\
\hline
  $\nu=2/3$   &  $N_s=36$, $N_b=6$     &   $N_s=48$, $N_b=8$   &    \\
\hline
  $\cal D$    &  $\sim 216421$      &   $\sim 31445749$  &     \\
\hline\hline
\end{tabular}
\caption{Dimensions ($\cal D$) of Hilbert space with periodic boundary conditions for $\nu=1/2$ and $\nu=2/3$ FCIs with various numbers of particles ($N_b$) and lattice sizes ($N_s$). Here, $\sim$ denotes approximately equal to the number.}
\label{tab1}
\end{table}

\bibliography{SQLattice}

\end{document}